\documentclass[
    ,final            
  ]
  {aipproc}

\def\nuc#1#2{${}^{#1}$#2}

\def\nonubb{$\beta\beta(0\nu)$}
\def\twonubb{$\beta\beta(2\nu)$}

\def\MJ{{\sc Majorana}}             
\def\be{\begin{equation}}
\def\ee{\end{equation}}

\def\DEMO{{\sc Demonstrator}}

\layoutstyle{6x9}

\begin{document}

\title{The {\sc Majorana} Experiment}


\newcommand{\alberta}{Centre for Particle Physics, University of Alberta, Edmonton, AB, Canada}
\newcommand{\blhill}{Department of Physics, Black Hills State University, Spearfish, SD, USA}
\newcommand{\ITEP}{Institute for Theoretical and Experimental Physics, Moscow, Russia}
\newcommand{\JINR}{Joint Institute for Nuclear Research, Dubna, Russia}
\newcommand{\lbnl}{Nuclear Science Division, Lawrence Berkeley National Laboratory, Berkeley, CA, USA}
\newcommand{\lanl}{Los Alamos National Laboratory, Los Alamos, NM, USA}
\newcommand{\queens}{Department of Physics, Queen's University, 
Kingston, ON, Canada}
\newcommand{\uw}{Center for Experimental Nuclear Physics and Astrophysics, 
and Department of Physics, University of Washington, Seattle, WA, USA}
\newcommand{\uchic}{Department of Physics, University of Chicago, Chicago, IL, USA}
\newcommand{\unc}{Department of Physics and Astronomy, University of North Carolina, Chapel Hill, NC, USA}
\newcommand{\duke}{Department of Physics, Duke University, Durham, NC, USA}
\newcommand{\ncsu}{Department of Physics, North Carolina State University, Raleigh, NC, USA}
\newcommand{\ornl}{Oak Ridge National Laboratory, Oak Ridge, TN, USA}
\newcommand{\ou}{Research Center for Nuclear Physics and Department of Physics, Osaka University, Ibaraki, Osaka, Japan}
\newcommand{\pnnl}{Pacific Northwest National Laboratory, Richland, WA, USA}
\newcommand{\sdsmt}{South Dakota School of Mines and Technology, Rapid City, SD, USA}
\newcommand{\usc}{Department of Physics and Astronomy, University of South Carolina, Columbia, SC, USA}
\newcommand{\usd}{Department of Earth Science and Physics, University of South Dakota, Vermillion, SD, USA}
\newcommand{\ut}{Department of Physics and Astronomy, University of Tennessee, Knoxville, TN, USA}
\newcommand{\tunl}{Triangle Universities Nuclear Laboratory, Durham, NC, USA}


\author{E.~Aguayo}{address={\pnnl}} 
\author{F.T.~Avignone~III}{address={\usc}, altaddress={\ornl}}
\author{H.O.~Back}{address={\ncsu, \tunl}} 
\author{A.S.~Barabash}{address={\ITEP}}
\author{M.~Bergevin}{address={\lbnl}} 
\author{F.E.~Bertrand}{address={\ornl}}
\author{M.~Boswell}{address={\lanl}} 
\author{V.~Brudanin}{address={\JINR}}
\author{M.~Busch}{address={\duke}, altaddress={\tunl}}	
\author{Y-D.~Chan}{address={\lbnl}}
\author{C.D.~Christofferson}{address={\sdsmt}} 
\author{J.I.~Collar}{address={\uchic}}
\author{D.C.~Combs}{address={\ncsu}, altaddress={\tunl}}  
\author{R.J.~Cooper}{address={\ornl}}
\author{J.A.~Detwiler}{address={\lbnl}}
\author{P.J.~Doe}{address={\uw}}
\author{Yu.~Efremenko}{address={\ut}}
\author{V.~Egorov}{address={\JINR}}
\author{H.~Ejiri}{address={\ou}}
\author{S.R.~Elliott}{address={\lanl}}
\author{J.~Esterline}{address={\duke}, altaddress={\tunl}}
\author{J.E.~Fast}{address={\pnnl}}
\author{N.~Fields}{address={\uchic}} 
\author{P.~Finnerty}{address={\unc}, altaddress={\tunl}}
\author{F.M.~Fraenkle}{address={\unc}, altaddress={\tunl}} 
\author{V.M.~Gehman}{address={\lanl}}
\author{G.K.~Giovanetti}{address={\unc}, altaddress={\tunl}}  
\author{M.P.~Green}{address={\unc}, altaddress={\tunl}}  
\author{V.E.~Guiseppe}{address={\usd}}	
\author{K.~Gusey}{address={\JINR}}
\author{A.L.~Hallin}{address={\alberta}}
\author{R.~Hazama}{address={\ou}}
\author{R.~Henning}{address={\unc}, altaddress={\tunl}}
\author{A.~Hime}{address={\lanl}}
\author{E.W.~Hoppe}{address={\pnnl}}
\author{M.~Horton}{address={\sdsmt}} 
\author{S. Howard}{address={\sdsmt}}  
\author{M.A.~Howe}{address={\unc}, altaddress={\tunl}}
\author{R.A.~Johnson}{address={\uw}} 
\author{K.J.~Keeter}{address={\blhill}}
\author{M.E.~Keillor}{address={\pnnl}}
\author{C.~Keller}{address={\usd}}
\author{J.D.~Kephart}{address={\pnnl}} 
\author{M.F.~Kidd}{address={\lanl}}	
\author{A. Knecht}{address={\uw}}	
\author{O.~Kochetov}{address={\JINR}}
\author{S.I.~Konovalov}{address={\ITEP}}
\author{R.T.~Kouzes}{address={\pnnl}}
\author{B.D.~LaFerriere}{address={\pnnl}}   
\author{B.H.~LaRoque}{address={\lanl}}	
\author{J. Leon}{address={\uw}}	
\author{L.E.~Leviner}{address={\ncsu}, altaddress={\tunl}}
\author{J.C.~Loach}{address={\lbnl}}	
\author{S.~MacMullin}{address={\unc}, altaddress={\tunl}}
\author{M.G.~Marino}{address={\uw}}
\author{R.D.~Martin}{address={\lbnl}}	
\author{D.-M.~Mei}{address={\usd}}
\author{J.H.~Merriman}{address={\pnnl}}   
\author{M.L.~Miller}{address={\uw}} 
\author{L.~Mizouni}{address={\usc}, altaddress={\pnnl}}  
\author{M.~Nomachi}{address={\ou}}
\author{J.L.~Orrell}{address={\pnnl}}
\author{N.R.~Overman}{address={\pnnl}}  
\author{D.G.~Phillips II}{address={\unc}, altaddress={\tunl}}  
\author{A.W.P.~Poon}{address={\lbnl}}
\author{G. Perumpilly}{address={\usd}}   
\author{G.~Prior}{address={\lbnl}} 
\author{D.C.~Radford}{address={\ornl}}
\author{K.~Rielage}{address={\lanl}}
\author{R.G.H.~Robertson}{address={\uw}}
\author{M.C.~Ronquest}{address={\lanl}}	
\author{A.G.~Schubert}{address={\uw}}
\author{T.~Shima}{address={\ou}}
\author{M.~Shirchenko}{address={\JINR}}
\author{K.J.~Snavely}{address={\unc}, altaddress={\tunl}}	
\author{V. Sobolev}{address={\sdsmt}}  
\author{D.~Steele}{address={\lanl}}	
\author{J.~Strain}{address={\unc}, altaddress={\tunl}}
\author{K.~Thomas}{address={\usd}}		
\author{V.~Timkin}{address={\JINR}}
\author{W.~Tornow}{address={\duke}, altaddress={\tunl}}
\author{I.~Vanyushin}{address={\ITEP}}
\author{R.L.~Varner}{address={\ornl}}  
\author{K.~Vetter}{address={\lbnl}, altaddress={Department of Nuclear Engineering, University of California, Berkeley, CA, USA}}
\author{K.~Vorren}{address={\unc}, altaddress={\tunl}} 
\author{J.F.~Wilkerson$^{\mathrm{j,}}$}{address={\unc}, altaddress={\tunl}, altaddress={\ornl}}    
\author{B.A. Wolfe}{address={\uw}}	
\author{E.~Yakushev}{address={\JINR}}
\author{A.R.~Young}{address={\ncsu}, altaddress={\tunl}}
\author{C.-H.~Yu}{address={\ornl}}
\author{V.~Yumatov}{address={\ITEP}}
\author{C.~Zhang}{address={\usd}}					

\begin{abstract}
The \MJ\ collaboration is actively pursuing research and development aimed at a tonne-scale \nuc{76}{Ge} neutrinoless double-beta decay (\nonubb -decay) experiment. The current, primary focus is the construction of the \MJ\ \DEMO\ experiment, an R\&D effort that will field approximately $40\,\mathrm{kg}$ of germanium detectors with mixed enrichment levels. This article provides a status update on the construction of the \DEMO .
\end{abstract}
\maketitle

\section{Introduction}
\label{se:Intro}
The \MJ\ collaboration~\cite{henn09, guis11} is actively pursuing research and development aimed at a tonne-scale \nuc{76}{Ge} neutrinoless double-beta decay (\nonubb -decay)~\cite{avi08} experiment. The current, primary focus is the construction of the \MJ\ \DEMO\ experiment, an R\&D effort that will field approximately $40\,\mathrm{kg}$ of germanium detectors with mixed enrichment levels. As a technical goal the collaboration hopes to demonstrate a background low enough to justify building a tonne-scale \nuc{76}{Ge} experiment. It also intends to test the claim of the discovery of the \nonubb-decay of \nuc{76}{Ge}~\cite{kla04a, kla06}.  \MJ\ is working collaboratively with the GERDA collaboration~\cite{sch05} to prepare for a single international tonne-scale \nuc{76}{Ge} experiment that combines the best technical features of the two experiments. GERDA is pursuing a novel liquid argon or nitrogen shield in which the germanium detectors are directly immersed, while \MJ\ is pursuing a more conservative compact shield consisting of high purity copper and lead.

\MJ\ and GERDA use High-Purity Germanium (HPGe) semiconductor diode detectors fabricated from \nuc{76}{Ge} enriched material. HPGe detectors have excellent energy resolution, which is crucial to reduce the backgrounds due to other types of radioactive decays in the detectors and surrounding materials. HPGe detectors are also intrinsically very clean and have negligible intrinsic radioactivity. Germanium-based experiments using HPGe detectors also have the best sensitivity to \nonubb-decay to date~\cite{baud99, aals02}. 

\section{The \MJ\ \DEMO\ }
\label{se:MJDEMO}

The \MJ\ \DEMO\ will consist of 40~kg of HPGe detectors. Of these, 20-30~kg will be enriched to 86\%~in \nuc{76}{Ge}. This is the minimum amount of material required to achieve the technical and scientific goals outlined in the Introduction. An important technical goal for the \DEMO\ is to demonstrate a background of 4~counts per tonne-year exposure in the \mbox{$4\,\mathrm{keV}$} region-of-interest (ROI) around the \mbox{$2039\,\mathrm{keV}$} $Q$-value of the decay after analysis cuts have been applied.

Cylindrical crystals of semiconducting HPGe are doped to make large ($\sim 500\,\mathrm{g}$) diodes. By applying a reverse biasing potential, typically a few kilovolts, to the diode, an electrical field is established inside the crystal. Ionizing radiation creates electron-hole pairs that drift under the influence of the internal electrical field and are collected at the electrodes of the crystal. Small current pulses are induced on the electrodes by these drifting charges, which can be detected using standard nuclear physics electronics.

The \DEMO\ will require ionizing radiation backgrounds a factor of order $100$ lower than what has been achieved with germanium detector technology to date~\cite{kla04b, aals02}. There are multiple sources of backgrounds. Ubiquitous, primordial U and Th, and their decay chain daughters are a significant concern and can contribute in several different ways. Gamma-rays from radioactive decays with sufficient energy to affect the ROI typically Compton scatter inside germanium with a scattering length of $\sim 1\,\mathrm{cm}$. This is a very different topology than a \nonubb -decay, which would deposit all the ionization energy from the two electrons in a $\sim 1\,\mathrm{mm}^3$ region in the crystal, making it highly localized. Most of \MJ 's analysis-based background reduction techniques rely on the ability to separate these multi-site events (MSE) from single-site events (SSE).

One of the dominant backgrounds is the Compton continuum of the $2.6\,\mathrm{MeV}$ gamma ray from \nuc{208}{Tl} decay from the \nuc{232}{Th} chain, which is primarily mitigated using electroformed copper as structural and shielding material. Another significant background is cosmogenic activation products in copper and germanium, specifically \nuc{68}{Ge} in germanium and \nuc{60}{Co} in copper and germanium. These cosmogenic backgrounds are mitigated with analysis cuts and by minimizing surface exposure times. Backgrounds from radon and prompt cosmic-rays are mitigated with liquid nitrogen boil-off and by running deep underground with a cosmic-ray veto, respectively. As stated, HPGe detectors have excellent energy resolution and the \twonubb -decay mode does not contribute a background. Other backgrounds, such as external gamma-rays and fission neutrons, are mitigated with shielding.

The \DEMO 's most important structural material is electroformed (e-formed) copper, which has been demonstrated to have extremely low activity~\cite{brod95}. E-formed copper is made by electroplating from ultra-pure copper nuggets onto stainless steel mandrels. Electroplating for \MJ 's copper is performed underground to remove cosmogenic \nuc{60}{Co}. The collaboration is currently operating an underground e-forming facility at the 4850~foot level of Sanford Underground Laboratory (SUL) in Lead, SD and at the shallow underground facility at Pacific Northwest National Laboratories in Richland, WA. The goal of these facilities is to achieve an early start of the slow process of copper electroplating in order to accelerate the overall schedule of the \DEMO . The current best limits for activity in \MJ\ e-formed copper is~$<0.7\,\mu \mathrm{Bq/kg}$~\cite{hopp09} for Th and~$<1.3\,\mu \mathrm{Bq/kg}$~\cite{hopp11} for U.

\nuc{76}{Ge} has a natural abundance of 7.44\% and detector germanium for \MJ\ must be enriched to 86\% or better. ISOFLEX USA is the vendor of the enriched material, and the enrichment is being performed at the Electro-Chemical Plant (ECP) in Zelonogorsk, Russia. The enriched germanium oxide will be transported from ECP inside a massive, steel-shielded container via ground transport and cargo ship to Oak Ridge, TN, where a commercial company will process the oxide into electronic grade material. The detector manufacturer will use standard zone-refining and Czochralski crystal growing techniques to further purify the material into detector grade material. The first batch of 20~kg enriched material is expected to arrive in Oak Ridge in September, 2011, with the remaining 15.5~kg the next year. Additional enriched material will also be provided by Russian \MJ\ collaborators.

The \DEMO\ is using P-type Point-Contact (PPC) detectors. These detectors are approximately $7\,\mathrm{cm}$ in diameter and $3\,\mathrm{cm}$ high. PPC detectors have a very small p-type contact, which has two benefits for \MJ . The first is that the capacitance of the detector is minimized, significantly reducing the series noise of the system compared to other HPGe detector designs. The second benefit is the sharp rise in the weighting fields very near the p-type contact, causing most of the induced signal to be generated only as the charge cloud approaches near the point-contact. This renders the detector's output pulse shape sensitive to the radial distribution of the initial charge deposits inside the crystal, since radially separated deposits will induce signals at different times. This is important for rejection of MSE vs. SSE in \MJ .

The detectors are arranged in seven five-crystal strings. Each string has the mounting fixtures for the crystals and front-end electronics. The front-end electronics consist of a JFET, amorphous germanium feedback resistor and connector traces mounted on a low-mass (63~mg) fused silica board. Stray capacitance between leads on the board provide the feedback capacitance. E-formed copper, Vespel\textregistered , and PTFE that were selected to be low background are the dominant structural materials of a string. Seven strings are mounted electronically and mechanically to an e-formed copper cold-plate, and the entire cold-plate and string assembly is mounted inside an e-formed copper cryostat and IR shield. Electrical and thermal connections are made from the cold-plate to the outside via an e-formed copper cryostat arm. A liquid nitrogen thermosyphon~\cite{bolo09} provides the cooling to maintain the crystals at 80~K. Outside the shield the cryostat arm is connected to a vacuum system and the dewar for the thermosyphon, as well as a break-out box that connects cabling from the front-end electronics to pre-amplifiers.

From the inside out the shield consists of 5~cm of e-formed copper, 5~cm of commercial OFHC-grade copper, 45~cm of commercial lead, 30~cm of polyethylene moderator, and an active plastic scintillator cosmic-ray veto. The entire inner lead and copper shield cavity is hermetically sealed and continuously purged with liquid nitrogen (LN) boil-off gas to reduce radon. Each cryostat is mounted with its vacuum system and thermosyphon to a part of the shield, called a monolith, which can be rolled in and out of the entire shield, easing installation. Each cryostat is surrounded by an e-formed copper track for placing and removing encapsulated calibration sources near the detectors and inside the shield. The cryostat and its internals will be assembled in a glovebox purged with LN boil-off and operated in a class~2000 cleanroom in the Davis campus on the 4850' level of SUL. Machine tools will also be located in this cleanroom.  
 
Electronic signals from the preamps are digitized using high resolution digitizers and the pulse-shapes saved for off-line analysis. The data acquisition and slow controls are controlled using the ORCA software package~\cite{howe04}. The energy of an event is computed using standard digital filters, and pulse-shape analysis can distinguish MSE Compton-scattered gamma rays and internal background decay backgrounds from SSE \nonubb -decays. Additional analysis cuts can remove background events that have energy deposits in multiple crystals. The collaboration has performed extensive Monte Carlo simulations of the \DEMO\ to verify these analysis cuts and to determine purity requirements for materials. Over \mbox{50,000} combinations of radioactive isotopes and detector components have been simulated to date and are compiled into a background model. \MJ\ collaborators have also developed detailed simulations of pulse-generation and charge drift inside the PPC detectors to characterize the electronic response of the detectors and help diagnose pathologies in pulse-shapes arising from events near the detector surfaces.

\MJ\ will probe physics other than \nonubb -decay. With a large target mass, low background and low thresholds, the \DEMO\ will also be able to search for low mass WIMPS, as demonstrated by CoGeNT~\cite{aals08, aals10}. The collaboration has deployed a low-background PPC detector at the Kimballton Underground Research Facility (KURF) in Virginia to study the low-energy background in PPC detectors. Because the \MJ\ detector technology does not distinguish between nuclear and electronic recoils, the \DEMO\ will also be sensitive to keV scale dark matter (aka. SuperWIMPS) that scatter off electrons. The \DEMO\ can also perform searches for axions and study low energy neutrino scattering from a neutrino source.

A prototype module of the \DEMO\ filled with natural germanium detectors will be deployed and operated at the surface at Los Alamos National Laboratory starting in the summer of 2012, with the first module of $12\,\mathrm{kg}$ enriched detectors coming on-line underground at SUL in the summer of 2013 and the second module of $18\,\mathrm{kg}$ enriched detectors online in spring of 2014. The \DEMO\ should test the discovery claim within a year or two of commissioning its first module.

\begin{theacknowledgments}
We acknowledge support from the Office of Nuclear Physics in the DOE Office of Science under grant 
numbers DE-AC02-05CH11231, DE-FG02-97ER41041, DE-FG02-97ER41033, DE-FG02-97ER4104, 
DE-FG02-97ER41042, DE-SCOO05054, DE-FG02-10ER41715, and DE-FG02-97ER41020. We acknowledge support 
from the Particle and Nuclear Astrophysics Program of the National Science Foundation through grant 
numbers PHY-0919270, PHY-1003940, 0855314, and 1003399. We gratefully acknowledge support from the 
Russian Federal Agency for Atomic Energy. We gratefully acknowledge the support of the U.S. Department 
of Energy through the LANL/LDRD Program. \hbox{N. Fields} is supported by the DOE/NNSA SSGF program. G.~K.~Giovanetti is supported by the DOE Office of Science Graduate Fellowship Program.
\end{theacknowledgments}

\bibliographystyle{aipproc}   
\bibliography{main}

\end{document}